# Design and processing as ultrathin films of a sublimable Iron(II) spin crossover material exhibiting efficient and fast light-induced spin transition

Miguel Gavara-Edo (ORCID: 0000-0002-0311-3874),[a] Francisco Javier Valverde-Muñoz (ORCID: 0000-0003-3578-5445),[a] M. Carmen Muñoz (ORCID: 0000-0003-2630-3897),[b] Safaa Elidrissi Moubtassim (ORCID: 0000-0002-6687-3854),[a] Francisco Marques-Moros (ORCID: 0000-0001-8199-5326),[a] Javier Herrero-Martín (ORCID: 0000-0003-1986-8128),[c] Kateryna Znovjyak (ORCID: 0000-0003-4818-8658),[d] Maksym Seredyuk orcid.org/0000-0003-2615-094X,[a,d]* José Antonio Real (ORCID: 0000-0002-2302-561X)[a]* and Eugenio Coronado (ORCID: 0000-0002-1848-8791)[a]*

[a]Instituto de Ciencia Molecular, Universidad de Valencia, Catedrático José Beltrán 2, Paterna 46980, Spain
[b]Departamento de Fisica Aplicada, Universitat Politècnica de València, Camino de Vera s/n, Valencia 46022, Spain
[c]CELLS-ALBA Synchrotron, Carrer de la Llum 2-26, Cerdanyola del Vallès 08290, Spain
[d]Department of Chemistry, Taras Shevchenko National University of Kyiv, 64/13, Volodymyrska Street, 01601 Kyiv, Ukraine

**Abstract**

Materials based on spin crossover (SCO) molecules have centred the attention in Molecular Magnetism for more than forty years as they provide unique examples of multifunctional and stimuli-responsive materials, which can be then integrated into electronic devices to exploit their molecular bistability. This process often requires the preparation of thermally stable SCO molecules that can sublime and remain intact in contact with surfaces. However, the number of robust sublimable SCO molecules is still very scarce. Here we report a novel example of this kind. It is based on a neutral iron (II) coordination complex formulated as [$Fe^{II}$(neoim)$_2$], where neoimH is the ionogenic ligand 2-(1*H*-imidazol-2-yl)-9-methyl-1,10-phenanthroline. In the first part a comprehensive study, which covers the synthesis and magneto-structural characterization of the [$Fe^{II}$(neoim)$_2$] complex as a bulk microcrystalline material, is reported. Then, in the second part we investigate the suitability of this material to form thin films through high vacuum (HV) sublimation. Finally, the retainment of all present SCO capabilities in the bulk when the material is processed is thoroughly studied by means of X-ray absorption spectroscopy. In particular, a very efficient and fast light-induced spin transition (LIESST effect) has been observed, even for ultrathin films of 15 nm.

**INTRODUCTION**

In hexacoordinated Fe$^{II}$ spin crossover (SCO) complexes, the energy gap between the so-called low-spin (LS, $t_{2g}^6 e_g^0$) and high-spin (HS, $t_{2g}^4 e_g^2$) states is of the order of magnitude of the thermal energy. This fact confers electronic lability to these materials[1–4] and, consequently, a LS↔HS spin switching can be triggered by applying various external stimuli such as temperature or pressure changes, light irradiation, electric fields and even by host-guest interactions.[5–9] Due to the antibonding nature of the $e_g$ orbitals, the spin transition from LS to HS is accompanied by changes in size and shape of the SCO active coordination centers, often leading to a significant expansion of the crystal lattice (by ca. 10%). Depending on the structural nature of the complex, these changes can spread over the crystals in a cooperative way leading to a thermal hysteresis and thus to memory effects, which are reflected in the magnetic, calorimetric, optical and even electrical properties of the material.[1–9] These unique characteristics have made Fe$^{II}$ SCO complexes a relevant class of molecule-based switchable materials with potential use as device components for applications in fields such as molecular electronics and spintronics. Still, to achieve this end, it is necessary to process these SCO materials at the nanoscale.[10–13]

Among the available strategies, High-vacuum (HV) sublimation has been demonstrated to be an excellent methodology to prepare (ultra)thin films of SCO molecules.[11] The most prolific family of sublimable Fe$^{II}$ SCO complexes is constituted of pyrazolylborate-based anionic ligands, such as hydro-trispyrazolylborate ([HB(pz)$_3$]$^-$) and its -3,5-dimethylpyrazolyl homologue ([HB(3.5-pz)$_3$]$^-$)[14,15] or, more recently, hydrotristriazolylborate ([HB(trz)$_3$]$^-$)[14,16] and dihydropyrazolylpyridylpirazoleborate ([H$_2$B(pz)(pzpy)]$^-$) (**Scheme I**).[17] The tridentate nature and negative charge of these particular ligands affords neutral homoleptic [Fe$^{II}$L$_2$] SCO complexes. In contrast, the use of the bidentate dihydrobispyrazolylborate ([H$_2$B(pz)$_2$]$^-$) ligand introduces a degree of freedom to the system generating a rich variety of [Fe$^{II}$(H$_2$B(pz)$_2$)$_2$(L')] heteroleptic SCO complexes, being L' chelate bidentate α-diimine ligands derived from either 1,10-phenanthroline or 2,2'-bipy[18–24] and other related such as tzpy, btz and bt (**Scheme I**).[25] Also, few additional SCO complexes have also been selected to assess their suitability as sublimable materials.[26–31]

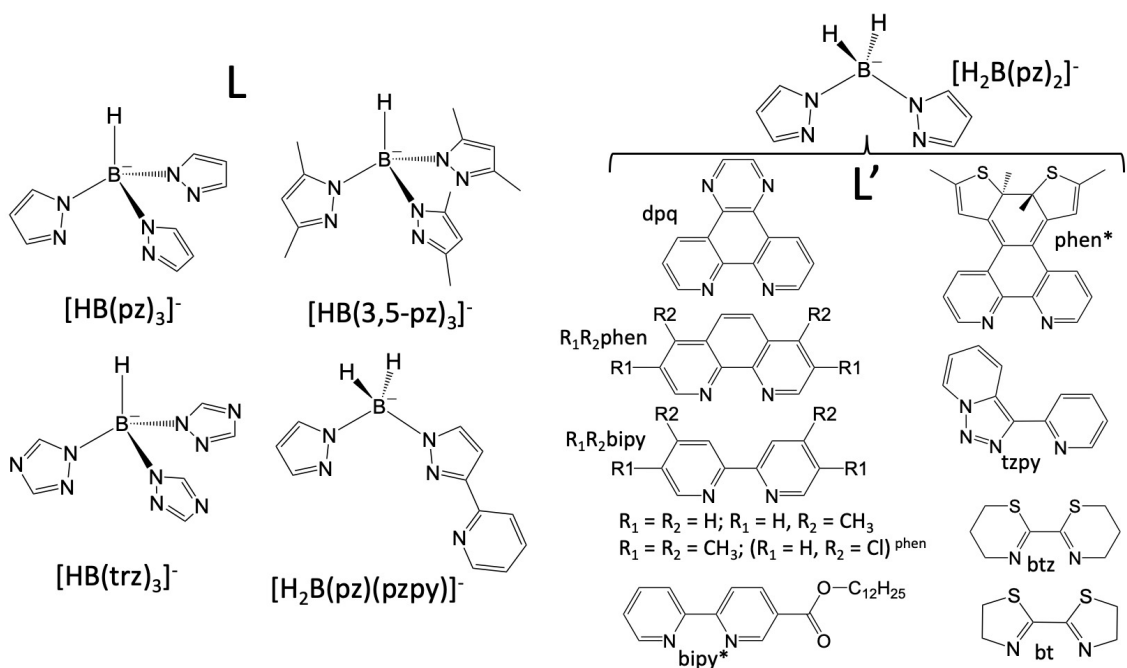

**Scheme I.** Tridentate and bidentate negatively charged ligands (L) of the pyrazolylborate-type and the ancillary chelate bidentate α-diimine ligands (L') used in combination with $[H_2B(pz)_2]^-$ one.

Studying the thermal stability of these compounds as well as the retainment of their switchable SCO properties once deposited as ultrathin films by sublimation or even as isolated molecules on different types of surfaces has become a growing research activity in the last 10 years,[21,23–25,32–42] as recently reviewed.[43,44] Here, we report the synthesis, processing as thin films and characterization of a novel neutral iron (II) coordination complex formulated as $[Fe^{II}(neoim)_2]$, where neoimH is the ionogenic ligand 2-(1*H*-imidazol-2-yl)-9-methyl-1,10-phenanthroline (see scheme 2). In bulk, this molecular complex exhibits both thermal and light-induced spin transitions. These properties are retained when this complex is deposited as thin films by sublimation, as shown by means of X-ray absorption spectroscopy, which also evidences the outstandingly fast and effective sensitivity of these films towards both visible and X-ray light-induced LS to HS conversions.

**RESULTS AND DISCUSSION**

**Synthesis**. The new neutral SCO complex $[Fe(neoim)_2]$ results from the design of ionogenic robust ligands based on the 1,10-phenanthroline since this function typically offers the appropriate ligand field strength to favor thermal spin-state conversion in the $Fe^{II}$ ion. To obtain this complex, we have functionalized the neocuproine ligand (2,9-dimethyl-1,10 phenanthroline) with a complementary imidazole function, thereby affording the novel tridentate ligand 2-(1*H*-imidazol-2-yl)-9-methyl-1,10-phenanthroline (hereafter neoimH). Then, upon deprotonation in presence of $Fe^{II}$ we have obtained [Fe(neoim)$_2$] (see **Scheme II** and Methods sections for details). This compound was first dried under Ar flow at 440 K during 12 h to afford an unsolvated green-brown colored powder. Upon recrystallization by slow diffusion of hexane vapors into a solution of the molecular complex in chloroform, dark-green single crystals of the solvate [Fe(neoim)$_2$]·H$_2$O·2CHCl$_3$ were next afforded. Nonetheless, collected IR spectra of both solvated and unsolvated forms were essentially identical, being the main difference the intense band at 746 cm$^{-1}$ which corresponds to the C-Cl stretching band of CHCl$_3$ solvent. Additionally, the loss of the solvents was observed to induce noticeable changes in the crystal packing of both forms. Regarding the thermal stability of the molecule, the thermogravimetric analysis showed that the unsolvated samples endure temperatures up to ca. 600 K, indicating large thermal stability and hinting its potential sublimability.

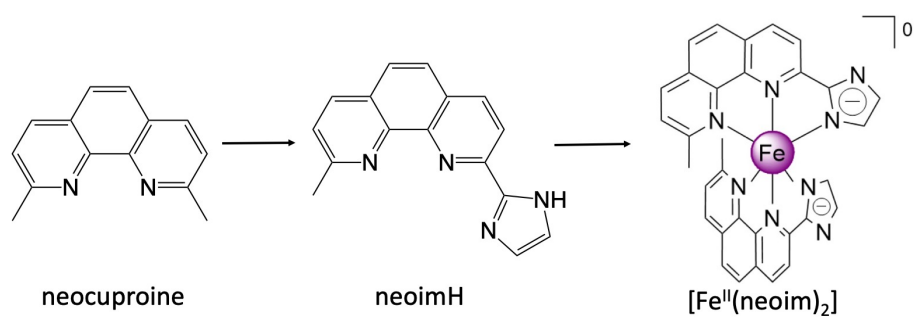

neocuproine            neoimH            [Fe$^{II}$(neoim)$_2$]

**Scheme II.** Simplified scheme of the pathway followed to afford the [Fe(neoim)$_2$] complex.

**Structure.** The crystal structure of [Fe(neoim)$_2$]·H$_2$O·2CHCl$_3$ was solved at 120 K. The unit cell corresponds to the chiral orthorhombic $P2_12_12_1$ space group. **Table 1** contains a selection of relevant Fe$^{II}$-N bond-lengths and angles. The Fe$^{II}$ center is at the center of a distorted [FeN$_6$] octahedral coordination environment generated by the two respective tridentate [neoim]$^-$ ligands. These are orthogonally oriented (89.89°) to each other around the Fe$^{II}$ center (see **Figure 1**). The average <Fe-N> bond length, equal to 1.969(8) Å, is characteristic of the LS state, however, the sum of the deviation from 90° of the 12 "*cis*" N-Fe-N angles, Σ = (|θ-90|) = 83.6° is significantly high but consistent with the geometric constraints induced by the tridentate nature of the ligand.

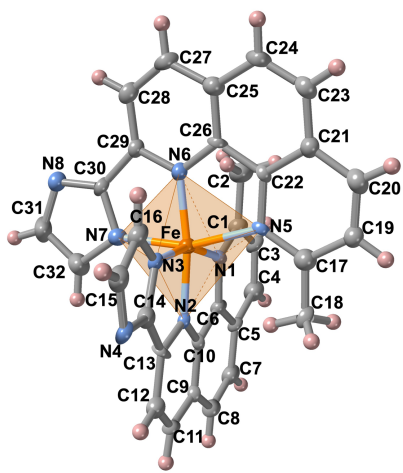

**Figure 1**. Molecular structure of compound [Fe(neoim)$_2$]·H$_2$O·2CHCl$_3$.

**Table 1**. Selected bond lengths (Å) and angles (°) of [Fe(neoim)$_2$]·H$_2$O·2CHCl$_3$.

| Bond lengths | | Bond angles | |
|---|---|---|---|
| Fe-N1 | 2.047(7) | N2 Fe N1 | 80.9(3) |
| Fe-N2 | 1.883(9) | N3 Fe N1 | 160.9(4) |
| Fe-N3 | 1.968(8) | N5 Fe N1 | 92.9(3) |
| Fe-N5 | 2.051(8) | N6 Fe N1 | 105.3(3) |
| Fe-N6 | 1.883(8) | N7 Fe N1 | 89.5(3) |
| Fe-N7 | 1.983(8) | N2 Fe N3 | 80.0(3) |
| <Fe-N> | 1.969(8) | N2 Fe N5 | 105.2(3) |
| | | N6 Fe N2 | 171.1 |
| | | N2 Fe N7 | 94.1(3) |
| | | N3 Fe N5 | 91.7(3) |
| | | N6 Fe N3 | 93.7(3) |
| | | N7 Fe N3 | 92.1(3) |
| | | N6 Fe N5 | 81.1(3) |
| | | N7 Fe N5 | 160.8(4) |
| | | N6 Fe N7 | 79.9(3) |
| | | Σ = (|θ-90|) | 83.6 |

Regarding the crystal packing of the complexes (**Figure 2**), a perfect superposition between them is found along *a*-direction defining columns. However, along *b* and *c* directions remarkably short intermolecular contacts are found. In particular, atoms C3 and C4 of the phenanthroline moiety of a molecule act as a wedge, filling the dihedral space generated between the imidazole and phenanthroline moieties of the adjacent complex and defining contacts well below the sum of the van der Waals radii [d(C3···C16) = 3.299 Å, d(C4···C21) = 3.193 Å] in addition to d(C31···C19) = 3.444 Å], which extend along *b* direction defining supramolecular chains. These chains are packed along *c* in such a way that short interchain contacts [d(C31···C24) = 3.437 Å, d(C31···C27) = 3.427 Å, d(C32···C27) = 3.362 Å] alternate with wide hexagonal channels running along *a* where the solvent molecules are located. The water molecule interacts via hydrogen bonding with one of the two chloroform molecules [d(C34···O1) = 3.092, angle C34-H···O1 = 174.67°] and also interacts with the C23 atom belonging to the phenanthroline moiety [d(O1···C7) = 3.371 Å]. In addition, the C11 of this moiety has a short contact with the Cl3 of the other chloroform molecule.

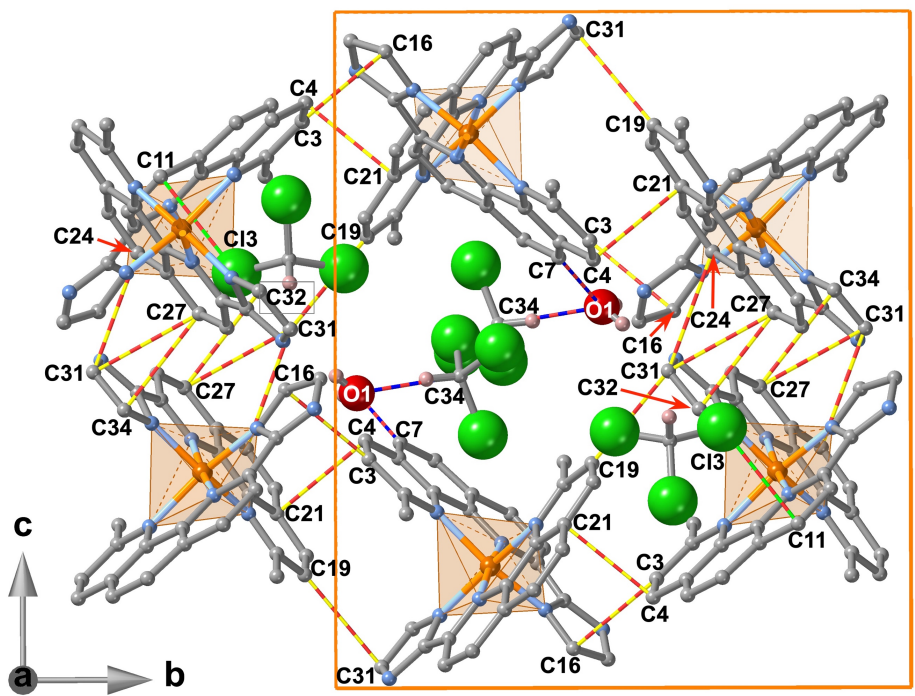

**Figure 2**. Representative crystal packing of [Fe(neoim)$_2$]·H$_2$O·2CHCl$_3$. Thin bicolor rods represent intermolecular short contacts. The hydrogen atoms of the complex have been omitted for clarity.

**Magnetic properties**. The magnetic properties of [Fe(neoim)$_2$]·H$_2$O·2CHCl$_3$ crystals and the corresponding desolvated form ([Fe(neoim)$_2$]) are displayed in **Figure 3** as the thermal dependence of the $\chi_M T$ product, where $\chi_M$ is the magnetic molar susceptibility and T is the temperature. For the solvated form, the $\chi_M T$ value is constant and around 0.1 cm$^3$ K mol$^{-1}$ in the whole temperature range, indicating that this material is essentially in the diamagnetic LS state. The loss of the solvent molecules provokes a dramatic change in the magnetic properties, which exhibit a spin transition. Thus, at 390 K, the $\chi_M T$ product shows a value ca. 3.49 cm$^3$ K mol$^{-1}$, consistent with a HS state. Upon cooling, $\chi_M T$ decreases very gradually down to ca. 1.28 cm$^3$ K mol$^{-1}$ at ca. 148 K. Upon further cooling, in the interval 148-132 K the slope of the $\chi_M T$ vs T plot increases significantly delineating an inclined short step followed by a steeper $\chi_M T$ vs T dependence until reaching a value ca. 0.30 cm$^3$ K mol$^{-1}$ at 80 K that stays constant down to the lowest temperature investigated and indicates that most of the Fe$^{II}$ centers are in a LS state. This second step exhibits a ca. 8 K wide hysteresis loop upon heating.

Photo-generation of the metastable HS* state from the LS state — the so-called light-induced excited spin state trapping effect (LIESST)[45] — was carried out at 10 K irradiating the desolvated powder with green light ($\lambda$ = 532 nm, 11.2 mW). Thus, $\chi_M T$ increases saturating to a value ca. 2.0 cm$^3$ K mol$^{-1}$ in 15 minutes, which considering the "high" temperature of the experiment and the penetration depth of the laser through the powder sample is noticeably fast and efficient. After switching off the light and heating the system at a rate of 0.3 K min$^{-1}$ a gradual increase of $\chi_M T$ is induced that attains a maximum value of 2.32 cm$^3$ K mol$^{-1}$ in the interval of 22-38 K. This small additional raise in $\chi_M T$ reflects the thermal population of different microstates originated from the zero-field splitting of the HS* state and corresponds to a population of the HS state ca. 66.5 %. Above 38 K, $\chi_M T$ decreases until joining the thermal SCO curve at ca. 80 K, indicating that the metastable HS* state has thermally relaxed back to the stable LS state. The corresponding $T_{LIESST}$ temperature, evaluated as $\delta(\chi_M T)/\delta T$,[2] is ca. 65 K. This temperature is consistent with the inverse-energy-gap law, i.e. the metastability of the photo-generated HS* species decreases as the stability of the LS increases.[46–49]

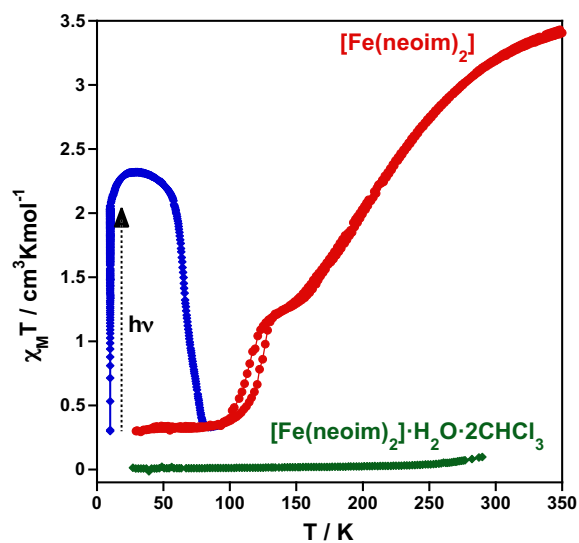

**Figure 3.** Magnetic properties of [Fe(neoim)$_2$]·H$_2$O·2CHCl$_3$ and its desolvated form.

**Highly homogeneous thin films of [Fe(neoim)$_2$]**

[Fe(neoim)$_2$] was sublimed under UHV conditions following a similar protocol used by us for other SCO molecules.[25,38] At this respect, ca. 100 mg of the desolvated bulk powder are loaded inside the Knudsen cell fitting our molecular beam epitaxy evaporation chamber (customized CREATEC system within a clean room class 10000) and preconditioned by means of a smooth heating up to 120°C for degassing. The optimal sublimation conditions consist in heating the material at 245°C in a vacuum pressure ca. $5·10^{-8}$ mbar to reach a constant deposition rate ca. 0.4 Å·s$^{-1}$ as monitored with a calibrated quartz crystal microbalance (QCM) located next to the deposition substrate. Different films were grown in the range 15 to 150 nm thick, verified with nanometric resolution via profilometry. The appearance and topography of these films were characterized using optical microscopy (OM) and atomic force microscopy (AFM). The results from both microscopies show extremely homogeneous coverages of the substrates (**Figure 4a,b**), featured by the lack of distinguishable crystal grains and by very low roughness (ca. 0.5 nm calculated as root mean square (RMS) value using Gwiddion program). No diffraction peaks can be observed by means of surface X-ray diffraction technique, indicating a highly amorphous character of the molecular films.

Regarding chemical integrity, the composition of the thin films was determined via IR and Raman spectroscopies and compared to that of the bulk (**Figure 4c,d**). Two 100 nm thick films were separately grown using the same sublimation procedure; one was deposited on a Au covered glass substrate destined for IR spectroscopy, and a second one was deposited on a $SiO_2$ substrate for Raman spectroscopy. The collected IR spectrum of this film shows a good matching with that obtained for the bulk powder between 2000 and 600 cm$^{-1}$ (**Figure 4c**). The only noticeable differentiating feature comes from the presence of a very small fraction of remnant water molecules within the bulk powder, evidenced by the presence of a noisy band at ca. 1600 cm$^{-1}$. Regarding the Raman spectra, the resulting data show a perfect matching between both film and bulk powder in the 1000 – 1700 cm$^{-1}$ region. Interestingly, no laser damaged was observed during this characterization on neither sample despite using a highly energetic laser (473 nm – blue, power ca. 1.08 mW·μm$^{-2}$). This result further evidences the good stability of this molecular SCO material. Overall, both techniques prove the retainment of the chemical integrity of the molecular complex when sublimed as thin film.

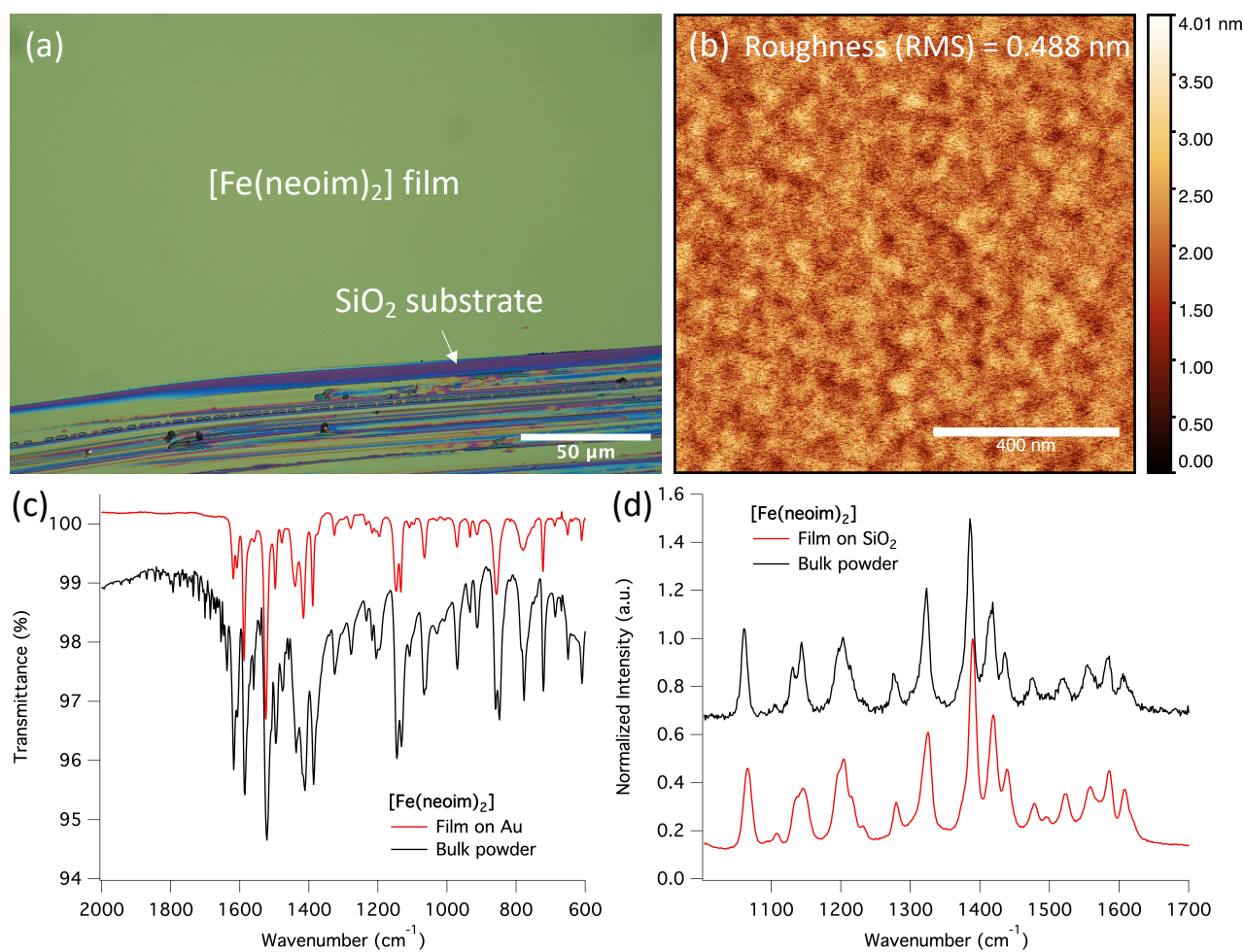

**Figure 4.** (a) OM image of a 100 nm thick [Fe(neoim)$_2$] film (green area) deposited on a SiO$_2$ substrate (purple area). (b) AFM image of a 1 μm x 1 μm region collected on the same film indicated in (a). (c) IR spectra of a 100 nm thick [Fe(neoim)$_2$] film deposited on a Au substrate (red line) and of the bulk powder reference (black line). (d) Raman spectra of a 100 nm thick [Fe(neoim)$_2$] film deposited on a SiO$_2$ substrate (red line) and of the bulk powder reference (black line). In (a) and (b) the scales are 50 μm and 400 nm respectively. In (b) the roughness of the film calculated as RMS value is indicated.

**Spin-crossover properties of the [Fe(neoim)$_2$] thin films**

The SCO behavior of different films was studied by means of X-ray absorption spectroscopy (XAS) technique focused at the Fe L$_{2,3}$ edges. Thus, XAS spectra were collected for each film at different temperatures to respectively determine the electronic configuration (spin state) of their conforming molecules. At the same time, the bulk material was characterized in order to get reliable referential spectra and to provide a direct correlation tool to estimate the HS fraction present in each sample at each temperature. Last but not least, the LIESST effect was additionally studied for the bulk material and the thinnest film prepared (15 nm thick) by irradiating each of them respectively at 2 K for 10 min with a red laser (633 nm, 12 mW) and subsequently collecting their respective XAS spectra. The results are presented in **Figure 5**.

To characterize the bulk material, spectra were collected at different temperatures by performing first a cooling from room temperature (300 K) to 2 K and then the subsequent heating up to 370 K. As it can be observed in (**Figure 5a,b**, red line), the initial state detected at 300 K can be assigned to a mixed HS-LS state where the HS one is predominant. This is characterized by the presence of both most characteristic peaks of the HS state at the Fe $L_3$ edge (ca. 708.0 and 708.9 eV respectively), as well as that of the LS state most intense peak (ca. 709.7 eV). Upon cooling down to 80 K, a clear progressive change is observed consisting on a decrease in intensity of the aforementioned HS state peaks at the Fe $L_3$ edge counterparted by an increase in intensity of the LS one (**Figure 5a**, black line). These observations, along the typical increase in intensity of the main peak of the Fe $L_2$ edge, appearing around 721 eV (**Figure 5b**, black line), are consistent with the thermal HS to LS conversion of the material characterized by magnetic susceptibility measurements. However, a deviation in such trend is observed upon further cooling. Particularly, the spectra collected down to 20 K show the progressive reversion of the spin state of the material towards a higher HS state fraction (**Figure 5a,b**, blue line). This behavior is typical of a soft X-ray induced excited spin state trapping (SOXIESST) effect that photoexcites the material from LS to a metastable HS* state during the spectra collection.[50] In fact, despite the precautions taken regarding the use of a very low photon flux, as performed in previous experiments on other sublimable SCO molecules,[25,38] the effect on the bulk [Fe(neoim)$_2$] molecule seems outstandingly effective. Thus, we next decided to cool further down to 2 K and irradiate with a red laser to investigate the LIESST effect as well with this technique. Interestingly, an almost full HS state spectrum is achieved only after 10 min irradiation (**Figure 5a,b**, orange line). Noticeably, apart from confirming the LIESST effect observed in magnetic susceptibility measurements, this result further evidences the high susceptibility of this material towards light-induced SCO phenomena. Subsequently, the reversibility of these observed SCO-related behaviors were studied upon heating up to 370 K. Two different trends can be distinguished: first, a thermal relaxation of both SOXIESST and LIESST effects is observed, with the full LS state recovered at ca. 100 K (**Figure 5a,b**, grey line); second, upon further heating, the expected progressive conversion from LS to an almost full HS state at 370 K occurs (**Figure 5a,b**, green line). In order to facilitate the visualization of all these changes, we estimated the temperature dependence of the HS fraction by fitting each collected spectrum to a linear combination of the "full" HS and LS states spectra (collected at 80 and 370 K, respectively) (**Figure 5c**). In this plot the different phenomenologies aforementioned are clearly monitored and quantified. For instance, at room temperature the bulk material presents a HS fraction of ca. 80 % which fully converts to the LS state at ca. 80 K; below this temperature the SOXIESST effect induces an overall HS fraction increase of ca. 20 % down to 20 K; finally, a complete LIESST effect is reached after 10 min red laser irradition at 2 K.

The XAS characterization of the sublimed thin films with three different thicknesess (100, 150 and 15 nm) is plotted in **Figure 5d-i**. The 100 nm film was submitted to a cooling between 300 K and 20 K followed by a heating to 240 K (**Figure 5d-f**). The results obtained closely resemble those attained for the bulk material, exhibiting a large HS fraction (ca. 95 %) at room temperature, which decreases down to ca. 37 % at ca. 80 K; at lower T clear SOXIESST effect is observed, reaching an outstandingly high HS fraction of ca. 68 % at 20 K (**Figure 5f**) despite using a very low photon flux as for the bulk powder (0.026 nA); and the completion of the SOXIESST effect's thermal relaxation is accomplished at 90 K upon heating.

Since the SOXIESST effect was very strong in this first sample and suspecting that the thermal SCO transition of the film might be more complete in absence of this phenomenon, a second sample with similar thickness (150 nm) was studied, but using a slightly lower photon flux (ca. 0.020 nA compared to 0.026 nA). Still, SOXIESST effect also emerged below 80 K (**Figure 5g-i**) and the overall behavior showed the same trend as for the first film but with noisier spectra. Consequently, the sample was not studied in further detail.

A last experiment was performed on a very thin film (15 nm) with the aim of investigating the effect of such miniaturization on the SCO behavior. Interestingly, the main SCO features are retained in this thin film (**Figure 5j-l**) including a SOXIESST effect below 80 K and an outstandingly effective LIESST effect. Due to the also very effective SOXIESST effect this film shows a more incomplete thermal SCO transition than the thicker ones. Although, the presence of a larger pinned fraction of molecules in the HS state in the 15 nm film could also contribute to this observation, given its larger surface to volume fraction.[51,52] As far as the LIESST effect is concerned, a full HS state is reached within the 10 min irradiation period. This remarkable result on this 15 nm film, together with the amorphous and highly homogeneous nature of the films of [Fe(neoim)$_2$], evidences the high suitability of the material to exploit its light-induced switchability in spintronic devices. In fact, one of the open goals in this area is that of fabricating multifunctional spin valve devices in which the transport properties can be not only tuned by applying a magnetic field but also by light irradiation. The reason why this goal has not been achieved so far is twofold. On the one hand, it is due to the difficulty to prepare homogeneous ultrathin films from SCO molecules fulfilling the requirements of a spintronic vertical device: spin transport through the film while, upon thinning, avoiding the presence of shorts and retaining the SCO features of the molecules. At this respect, so far, only few vertical <u>electronic</u> devices have been fabricated mostly integrating thin layers of SCO molecules belonging to the [Fe(H$_2$B(pz)$_2$)$_2$(L')] family,[53–55] which tipically display LIESST effect, grow as homogeneous and amorphous films and whose first layer of molecules is known to sometimes decompose on metal surfaces of conventional electrodes.[23–25,33,56–63] Moreover, these devices have generally presented high limitations for electrically reading out the SCO behaviors due to the strongly insulating character of the molecules and, in relation to this, have presented low ciclability given the high voltages required and the molecules reactivity.[53–55] On the other hand, the further limitation that has been preventing the use of LIESST in spintronic devices is that it has only been reported one example of a sublimable SCO molecule exhibiting an efficient and effective LIESST effect electrical response within an integrated electronic device.[38] Accordingly it acomplishes a full photoexcitation in minutes yet, unfortunately, the grown films of this material present high roughness and are composed by crystallites of ca. 50 nm, thus preventing its integration as ultra-thin films in vertical devices.

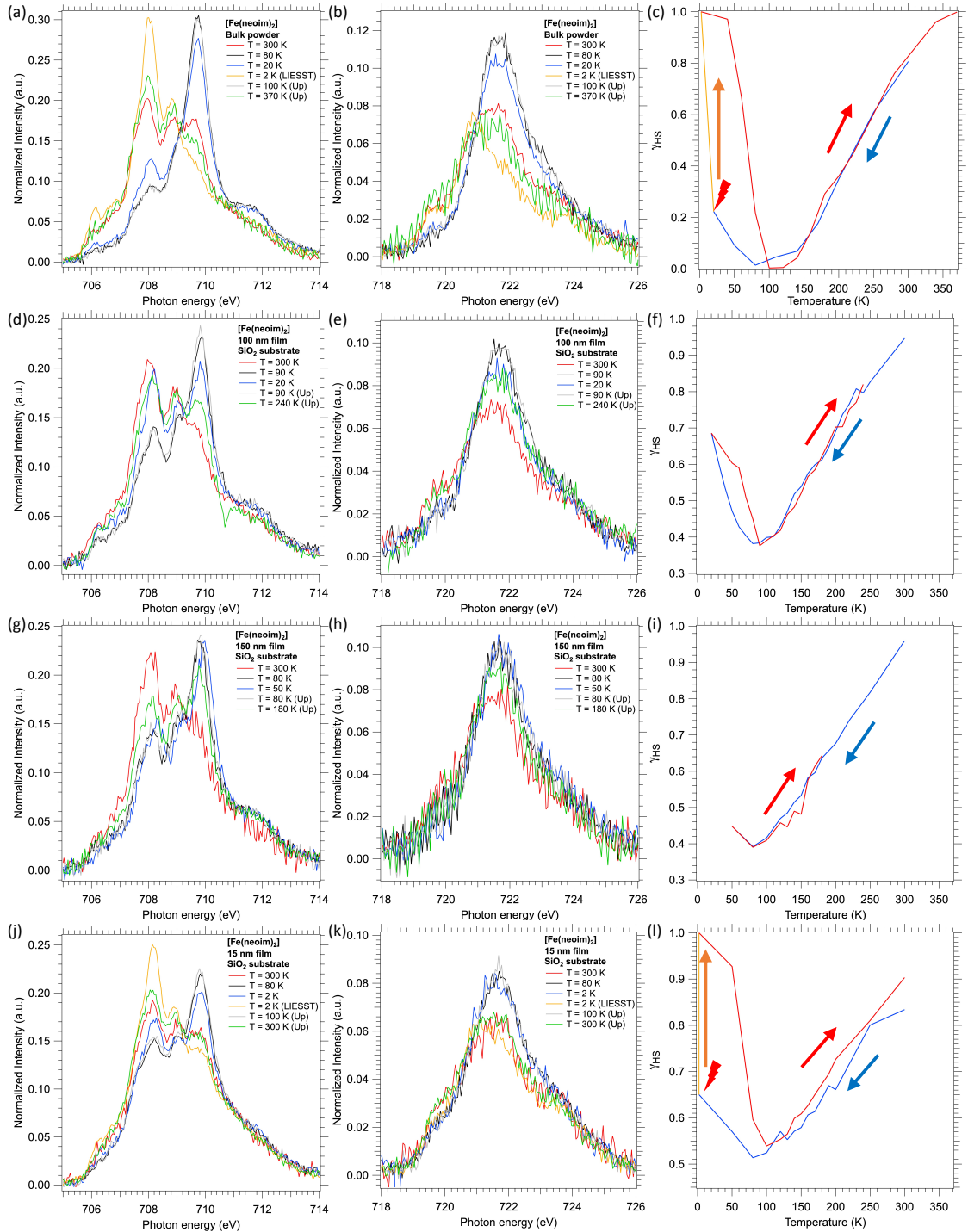

**Figure 5.** XAS spectra collected at 300, 80 and 20 K during cooling, 2 K after 10 min red laser irradiation, and 100 and 370 K during heating in the Fe a) L$_3$ and b) L$_2$ edges for [Fe(neoim)$_2$] scattered on C-tape and c) calculated HS fraction from each XAS spectra collected at each temperature along the full thermal cycle (blue line – cooling, orange line – LIESST and red line – heating). XAS spectra collected at 300, 90 and 20 K during cooling, and 90 and 240 K during heating in the Fe d) L$_3$ and e) L$_2$ edges for a 100 nm thick film of [Fe(neoim)$_2$] deposited on SiO$_2$ and f) calculated HS fraction from each XAS spectra collected at each temperature along the full thermal cycle (blue line – cooling and red line – heating). XAS spectra collected at 300, 80 and 50 K during cooling, and 80 and 180 K during heating in the Fe g) L$_3$ and h) L$_2$ edges for a 150 nm thick film of [Fe(neoim)$_2$] deposited on SiO$_2$ and i) calculated HS fraction from each XAS spectra collected at each temperature along the full thermal cycle (blue line – cooling and red line – heating). XAS spectra collected at 300, 80 and 2 K during cooling, 2 K after 10 min red laser irradiation, and 100 and 300 K during heating in the Fe j) $L_3$ and k) $L_2$ edges for a 15 nm thick film of [Fe(neoim)$_2$] deposited on SiO$_2$ and l) calculated HS fraction from each XAS spectra collected at each temperature along the full thermal cycle (blue line – cooling, orange line – LIESST and red line – heating).

**CONCLUSIONS**

Herein, we have reported a novel example of sublimable SCO material exhibiting extraordinary light-induced spin transition properties, which have been preserved when going from the bulk material to the thin films. This robustness has been accomplished departing from the chemical design of a novel ionogenic tridentate ligand (neoim) that affords the new neutral complex [Fe(neoim)$_2$]. The desolvated form of this complex displays a very progressive thermal SCO transition ranging from 80 K to 390 K, but, more importantly, it experiences a quantitative and fast light-induced spin transition (LIESST effect). The sublimation of this material yields the formation of remarkably homogeneous and amorphous thin films in the range 15 to 150 nm whose chemical integrity is identical to that of the bulk material. The SCO properties of these films have been determined by XAS measurements. These studies indicate that, in terms of both thermal and light-induced spin transitions, the behavior of these films matches very well with that of the bulk powder. Thus, the thermal SCO behavior of the bulk material is to a large extent well-preserved in the films. More importantly, the LIESST effect is well found to occur in both bulk and thin films. Remarkably, the retainment of such phenomenon in the ultrathin films makes this material highly appealing for its implementation in applications such as electronic and spintronic devices. In fact, such phenomenon has been so far poorly exploited in the field of molecular electronics,[38,53,64] since not many SCO materials display it, especially in the case of sublimable ones, and their performance is generally not very outstanding in terms of quickness and effectiveness. In spintronic devices such phenomenon has not been still exploited. Hence, in a next step, this material could be envisioned for its implementation in horizontal electronic devices integrating these SCO molecules with conducting/semiconducting 2D materials,[25,38,64–69] as well as in vertical spintronic devices in which the SCO ultrathin film is encapsulated in between two ferromagnetic electrodes. More specifically, we aim to improve the few reported results attained so far by exploiting the light-induced SCO switchability of this type of magnetic molecular materials.

**ACKNOWLEDGEMENTS**

The authors acknowledge the financial support from the European Union (ERC AdG Mol-2D 788222 and FET OPEN SINFONIA 964396), the Spanish MICINN (2D-HETEROS PID2020-117152RB-100, co-financed by FEDER, SPINCROSMAT PID2019-106147GB-I00 funded by MCIN/AEI/10.13039/501100011033 and Excellence Unit "María de Maeztu", CEX2019-000919-M), the Generalitat Valenciana (Prometeo program PROMETEO/2021/022) and Ministry of Education and Science of Ukraine (grants 22BF037-03 and 22BF037-04). M.G.-E. acknowledges the support of a fellowship FPU15/01474 from MIU. F.J.V.-M. acknowledges the support of the Generalitat Valenciana (APOSTD/2021/359). S.E.M. acknowledges the support of the Generalitat Valenciana with a Santiago Grisolía fellowship (GRISOLIAP/2018/046). All XAS experiments were performed at Boreas beamline at ALBA Synchrotron with J.H.-M. in both proposal and in-house experiments. The authors thank Alejandra Soriano Portillo and Ángel López Muñoz for their technical support.

**METHODS**

**Synthesis**

Iron(II) tetrafluoroborate hexahydrate, $MnO_2$, glyoxal, $NH_3$, and solvents were obtained from commercial sources and used as received without further purification. The sequence for synthesizing the ligand neoim and the corresponding $Fe^{II}$ complex involves the successive synthesis of 9-methyl-1,10-phenanthroline-2-carbaldehyde (**B**), the ligand neoimH (**C**) and the complex [Fe(neoim)$_2$] (**D**) as depicted in the following scheme:

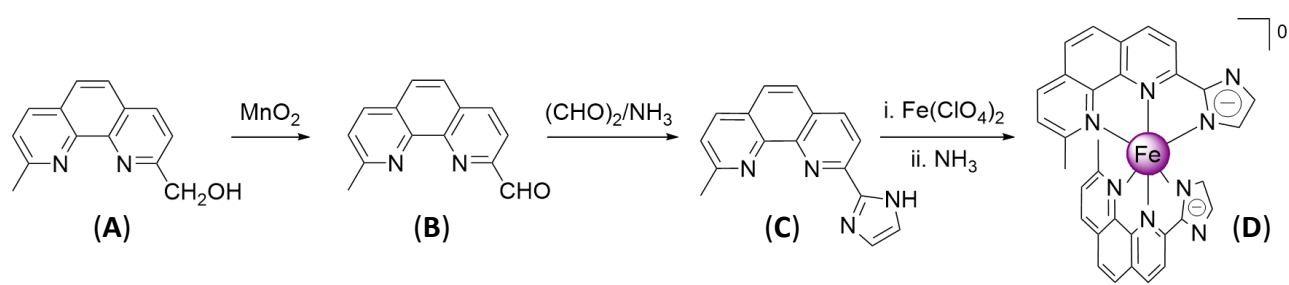

**9-methyl-1,10-phenanthroline-2-carbaldehyde (B).** To (9-methyl-1,10-phenanthrolin-2-yl)methanol (**A**) (10 g, 44 mmol)[70] dissolved in i-PrOH (200 ml), active $MnO_2$ (20 g) was added. The reaction mixture was refluxed under stirring for 2h and then, after cooling, filtered through a layer of celite. The inorganic $MnO_2$ cake was twice washed by boiling i-PrOH (in portions of 50 ml each). The organic solution was evaporated to dryness. The resulting yellow-brown residue was purified by column chromatography on silica gel, using $CHCl_3$ as the eluent. The intense white opaque band was collected. The yield of the target compound is about 5.0 g (51%). $^1$H NMR (300 MHz, CDCl$_3$): δ 10.47 (1H, s, CHO), 8.32 (1H, dd, $J^1$ = 0.7 Hz, $J^2$ = 8.3 Hz, CH$^7$), 8.19 (1H, d, $J$ = 8.2 Hz, CH$^4$), 8.11 (1H, d, $J$ = 8.3 Hz, CH$^8$), 7.84 (1H, d, $J$ = 8.8 Hz, CH$^5$), 7.73 (1H, d, $J$ = 8.8 Hz, CH$^6$), 7.52 (1H, d, $J$ = 8.2 Hz, CH$^3$), 2.92 (3H, s, Me). $^{13}$C NMR (100 MHz, CDCl$_3$): δ 193.99, 160.50, 152.04, 145.70, 145.48, 137.48, 136.58, 131.29, 129.40, 127.13, 125.24, 124.43, 119.52, 25.94. Anal. Calcd for C$_{14}$H$_{10}$N$_2$O: C, 75.66; H, 4.54; N, 12.60. Found: C, 75.78; H, 4.35; N, 12.57.

**2-(1*H*-imidazol-2-yl)-9-methyl-1,10-phenanthroline (neoimH) (C).** To a solution of 9-methyl-1,10-phenanthroline-2-carbaldehyde (5.0 g, 22 mmol) in EtOH (10 ml), aqueos glyoxal, 40% (3.3 ml, 29 mmol), was added and the mixture was cooled to 0 °C. To the obtained solution, NH$_3$(aq), 25% (9.0 ml, 132 mmol), was added dropwise with constant stirring during 4 h. The mixture was brought to room temperature and left overnight to produce an orange-brown solution. The volatiles were removed on rotary evaporator, the obtained solid dissolved in CHCl$_3$ (50 ml), dried over MgSO$_4$, filtered and purified by column chromatography on silica gel, using CHCl$_3$ as the eluent. The intense yellow opaque band was collected. The yield of the oven dried (8h, 80 °C) ligand is 2.3 g (39 %). $^1$H NMR (300 MHz, DMSO-d$^6$): δ 13.40 (1H, s br, NH), 8.52 (1H, d, $J$ = 8.5 Hz, CH$^7$), 8.41 (1H, d, $J$ = 8.5 Hz, CH$^8$), 8.39 (1H, d, $J$ = 8.2 Hz, CH$^4$), 7.93 (2H, s br, CH$^{5,6}$), 7.68 (1H, d, $J$ = 8.2 Hz, CH$^3$), 7.30 (2H, s br, imH), 2.84 (3H, s, CH$_3$). $^{13}$C NMR (100 MHz, CDCl$_3$): δ 159.20, 148.85, 146.81, 144.77, 137.78, 137.44, 128.30, 127.46, 126.62, 126.10, 124.60, 119.85, 56.51, 24.99, 18.98. Anal. Calcd for C$_{16}$H$_{12}$N$_4$: C, 73.83; H, 4.65; N, 21.52. Found: C, 73.58; H, 4.87; N, 21.47.

**Complex [Fe(neoim)$_2$] (D).** To a solution of neoimH (1000 mg, 3.8 mmol) in MeOH (20 ml), Fe(BF$_4$)$_2$·6H$_2$O (640 mg, 1.9 mmol) was added. The resulting dark red solution was refrigerated (4°C) overnight, and the plate-like red crystals that formed were filtered off, air-dried, and suspended in a mixture of NH$_3$(aq), 25% (30 ml), and CHCl$_3$ (100 ml). The violet-colored organic layer was separated, and the aqueous phase was extracted three more times with CHCl$_3$ (in portions of 50 ml each). The organic solutions were combined, dried over MgSO$_4$, and evaporated to dryness, producing a brown-colored powder of the complex. Yield is 910 mg, 79 %. Anal. Calcd for C$_{34}$H$_{28}$FeN$_8$: C, 67.56; H, 4.67; N, 18.54. Found: C, 67.25; H, 4.93; N, 18.18.

**Complex [Fe$^{II}$(neoim)$_2$]·H$_2$O·2CHCl$_3$.** Aiming at charactering the neutral complexes through X-ray diffraction studies, single crystals of the H$_2$O-CHCl$_3$ solvate was obtained by slow diffusion of hexane vapors into solutions of [Fe$^{II}$(neoim)$_2$] microcrystalline samples (100 mg) in boiling chloroform (7 ml). The [Fe$^{II}$(neoim)$_2$]·H$_2$O·2CHCl$_3$ compound was obtained as dark-green block single crystals. Anal. Calcd for C$_{36}$H$_{32}$Cl$_6$FeN$_8$O: C, 50.20; H, 3.75; N, 13.01. Found: C, 49.88; H, 3.98; N, 13.44.

**Physical characterization**

*Magnetic measurements* were performed on crystalline samples (20-40 mg) with a Quantum Design MPMS-XL-5 SQUID magnetometer working in the 2 to 400 K temperature range with an applied magnetic field 1 T. Experimental susceptibilities were corrected for diamagnetism of the constituent atoms by the use of Pascal's constants.

*Infra-red spectra*. The solid-state absorption IR spectrum was recorded with an Agilent Technologies Cary 630-FTIR spectrometer equipped with a diamond micro-ATR accessory in the 4000–400 cm$^{-1}$ range.

*X-ray Powder diffraction (XRPD)* measurements were performed on a PANalytical Empyrean X-ray powder diffractometer (monochromatic CuKα radiation) in capillary measurement mode.

*Thermogravimetric analysis* was performed on a Mettler Toledo TGA/SDTA 851e, in the 290–800 K temperature range under a nitrogen atmosphere with a rate of 10 K min$^{-1}$.

*Elemental analyses* (C, H, N) were performed with a CE Instruments EA 1110 CHNS Elemental analyzer.

*Single crystal X-ray measurements*. Single crystals were mounted on a glass fiber using a viscous hydrocarbon oil to coat the crystal and then transferred directly to the cold nitrogen stream for data collection. X-ray data were collected on a Supernova diffractometer equipped with a graphite monochromated Enhance (Mo) X-ray Source (λ = 0.71073 Å). The program CrysAlisPro, Oxford Diffraction Ltd., was used for unit cell determinations and data reduction. Empirical absorption correction was performed using spherical harmonics, implemented in the SCALE3 ABSPACK scaling algorithm. The structures were solved by direct methods using SHELXS-2014 and refined by full matrix least-squares on $F^2$ using SHELXL-2014.[71] Non-hydrogen atoms were refined anisotropically, and hydrogen atoms were placed in calculated positions refined using idealized geometries (riding model) and assigned fixed isotropic displacement parameters. CCDC file number 2260043 contains the supplementary crystallographic data for this paper. These data can be obtained free of charge from The Cambridge Crystallographic Data Centre via www.ccdc.cam.ac.uk/data_request/cif.

*Sublimation of thin films.* Thin films were sublimed under HV by heating the desolvated bulk powder of [Fe(neoim)$_2$] molecule, priorly submitted to a degassing protocol (120 °C under 10$^{-8}$ mbar pressure), in the conditions indicated above in the results description within a customized CREATEC Molecular Beam Epitaxy system placed in a Clean Room Class 10000. All prepared films had their thickness verified by means of profilometry using KLA Alpha-Step D-500 instrument with nanometric resolution.

*Optical microscopy*. OM imaging was performed using a Nikon Eclipse LV-150N microscope coupled to a Nikon DS-FI3 camera and through a 50x objective.

*Atomic Force Microscopy.* AFM images were collected using a Bruker Dimension Icon with Scan Assyst in tapping mode and processed and analyzed using Gwyddion program. The statistical value of roughness was calculated using this software as RMS value.

*Infrared spectroscopy.* IR spectrum was collected using a Fourier Transformation-Infrared Spectrometer NICOLET 5700 from Thermo Electron Corporation equipped with a module that allows measuring the transmittance of the reflected IR light from a film sample between 3100 cm$^{-1}$ and 650 cm$^{-1}$ and specifically grown on a 3 cm × 3 cm Au coated glass substrate.

*Raman spectroscopy.* Raman spectra were collected separately on a film grown on a $SiO_2$ substrate and on the bulk powder directly scattered on a glass-slide using the same equipment and acquisition parameters, Horiba LabRAM HR Evolution equipped with a 473 nm Laser beam with a maximum power of 1.08 mW·μm$^{-2}$ and in the 1000 – 1700 cm$^{-1}$ Raman shift region.

*X-ray absorption spectroscopy*. XAS characterization on the different films of [Fe(neoim)$_2$] indicated in the results discussion (deposited on 7 mm × 3 mm Si/SiO$_2$ substrates) and the desolvated bulk powder was performed at Boreas beamline in ALBA synchrotron during both granted beamtimes and in-house experiments. The substrates films were fixed to copper sample holders using aluminum clips and sitting on pieces of indium foil for their proper thermalization. Bulk powder was directly scattered onto C-tape stripes attached to the copper sample holder. For each case, the measurements implied the collection of three consecutive energy scans (for averaging) at each of a series of different temperatures within the range 2 – 370 K in separately cooling and heating modes and focused at the Fe $L_{2,3}$ edge region. The scans were performed using total electron yield mode with a low photon flux (0.020 nA ≤ intensity ≤ 0.025 nA) and only exposing the samples to X-rays during the scan collection periods. For the LIESST effect studies, the samples were irradiated with a red laser (He-Ne LASER from Research Electro-optics Inc. (R-30993), wavelength 633 nm and power 12 mW) at 2 K, right after the cooling process, and for only 10 min. All collected spectra for this study were processed for their analysis through background subtraction and normalization using Igor program. The dependence of the HS fraction (%) with temperature for each case was calculated from a fitting to a linear combination of the spectra showing the characteristic shapes for full and none (approximately) HS fractions of the bulk material (370 K during heating and 80 K during cooling respectively).